\documentclass[letterpaper]{article}
\usepackage{graphicx}
\usepackage{amsmath}
\usepackage{amssymb}
\usepackage{fancyheadings}
\usepackage{epsfig}
\usepackage{pstricks,pst-node}

\newlength{\figurewidth}
\newlength{\enviropost}
\newcommand{\be}{\begin{equation}}
\newcommand{\ee}{\end{equation}}
\newcommand{\ble}[1]{\begin{equation} \label{#1}}
\newcommand{\bae}{\begin{eqnarray}}
\newcommand{\eae}{\end{eqnarray}}
\newcommand{\fle}[2]%
{\vspace{1.5ex}
\be
\label{#1}
\mbox{%
\setlength{\fboxsep}{3ex}%
\framebox{$\dss #2 $}}
\ee} 
\newcommand{\flec}[2]%
{\vspace{1.5ex}
\be
\label{#1}
\mbox{%
\setlength{\fboxsep}{3ex}%
\framebox{$\dss #2 $}}
\, \, \,  ,
\ee} 
\newcommand{\flep}[2]%
{\vspace{1.5ex}
\be
\label{#1}
\mbox{%
\setlength{\fboxsep}{3ex}%
\framebox{$\dss #2 $}}
\, \, \, .
\ee} 

\newcommand{\nn}{\nonumber}
\newcommand{\ff}{\nn \\}
\newcommand{\fe}{& = &}

\newcommand{\dss}{\displaystyle}

\newcommand{\ket}[1]{| #1 \rangle}
\newcommand{\bra}[1]{\langle #1 |}

\newcommand{\eg}{\hbox{\em e.g.{}}}

\newcommand{\capitem}[1]{\caption{\textsf{#1}}}

\newcommand{\calO}{\mathcal{O}}

\newcommand{\papertitle}{%
Wires with Quantum Memory%
}
\newcommand{\runningtitle}{%
Wires with Quantum Memory%
}
\newcommand{\paperauthor}{%
C.{} Chryssomalakos, H.{} Hernandez, D.{} Gelbwaser-Klimovsky and E.{} Okon%
}
\begin{document}
\begin{titlepage}
\vspace*{-1cm}
\begin{flushright}
\textsf{}
\\
\mbox{}
\\
\textsf{\today}
\\[3cm]
\end{flushright}
\renewcommand{\thefootnote}{\fnsymbol{footnote}}
\begin{LARGE}
\bfseries{\sffamily \papertitle}
\end{LARGE}

\noindent \rule{\textwidth}{.6mm}

\vspace*{1.6cm}

\noindent \begin{large}%
\textsf{\bfseries%
\paperauthor
}
\end{large}


\phantom{XX}
\begin{minipage}{.8\textwidth}
\begin{it}
\noindent Instituto de Ciencias Nucleares \\
Universidad Nacional Aut\'onoma de M\'exico\\
Apdo. Postal 70-543, 04510 M\'exico, D.F., M\'EXICO \\
\end{it}
\texttt{chryss, hcoronado, david.gelbwaser, eliokon@nucleares.unam.mx
\phantom{X}}
\end{minipage}
\\

\vspace*{3cm}
\noindent
\textsc{\large Abstract: }
We show that quantum particles constrained to move along curves undergoing cyclic deformations acquire, in general, geometric phases. We treat explicitly an example, involving particular deformations of a circle, and ponder on potential applications.\end{titlepage}
\setcounter{footnote}{1}
\renewcommand{\thefootnote}{\arabic{footnote}}
\setcounter{page}{2}




\section{Introduction}
You leave a wire on a table in a locked room, and rumours soon have it that somebody entered the room, played with the wire, and left it at the exact same initial position --- is it possible to detect the intrusion?

Geometrical phases may appear in quantum systems the hamiltonian $H$ of which depends on a set of external parameters $\xi^A$, $H=H(\xi)$. After a cyclic change of the parameters, along a loop $C$ in parameter space, slow enough to guarantee that the system remains in an instantaneous eigenstate of $H$, the wavefunction acquires a phase that, apart from the standard dynamical part of the ``energy times time'' type, contains a contribution that only depends on $C$ --- hence the term ``geometric''~\cite{Ber:84}. Observable effects are obtained if the system starts out in a suitable superposition of energy eigenstates, each of which acquires a different geometric phase --- numerous experiments have confirmed these theoretical predictions (see, \eg, ~\cite{Tom.Chi:86,Bit.Dub:87,Sut.Chi.Har.Pin:87,Tyc:87,Zwa.Koe.Pin:90,%
Sut.Mue.Pin:88}). 

On a different vein, the dynamics of quantum particles constrained to move along a curve in three dimensional euclidean space, by a steep confining potential in the plane normal to the curve, is governed by an effective hamiltonian that depends on the curvature $\kappa(s)$ and the torsion $\tau(s)$ of the curve, where $s$ denotes arclength~\cite{Jen.Kop:71,Cos:81,Tak.Tan:92,Mar.Des:93} (see also~\cite{Mar:95,Mar:96,Fuj.Oga.Uch.Che:97,Sch.Jaf:03,Mit:01}). Thus, the Fourier coefficients, for example, of $\kappa$ and $\tau$, can be considered as external parameters of the effective hamiltonian. Slow, cyclic changes in the shape of the curve may then lead to the appearance of geometric phases, and it is the confirmation of this possibility that we report on here. Thus, the opening question is answered to the affirmative, at least for a playful enough intruder (see below).
Apart from limiting what intruders can get away with, the effect pointed out here opens up a wide arena for experimentation, both {\em gedanken} and real --- we ponder on some possibilities in the concluding section.
\section{Geometric phases}
Consider a hamiltonian $H_\xi$, as above, where the $\xi$'s are varying with time, tracing out a loop $C$ in $\xi$-space. Suppose the physical system governed by $H_\xi$ starts out, at time $t=0$, in the nondegenerate eigenstate $\ket{n,\xi_0}$, which satisfies $H_{\xi_0}\ket{n,\xi_0}=E_n^{\xi_0}\ket{n,\xi_0}$. Then, in the adiabatic approximation, in which the change in $H$ is slow (in the time scale set by the energy difference of neighboring eigenstates), the system's state at time $t$ is
\begin{equation}
\label{psit}
\ket{n,t}=e^{-i \alpha_n(t)+i\gamma_n(t)}\ket{n,\xi_t}
\, ,
\end{equation}
where $\ket{n,\xi_t}$ is an instantaneous eigenket of the hamiltonian, $H_{\xi_t}\ket{n,\xi_t}= E_n^{\xi_t}\ket{n,\xi_t}$, the phase $\alpha_n(t)=\int_0^t d\tau \, E_n^{\xi_\tau}$ is the expected dynamical one, and 
\begin{equation}
\label{gammandef}
\gamma_n(t)=i\int_0^t d\tau \, \bra{n,\xi_\tau} \frac{d}{d\tau} \ket{n,\xi_\tau}=i\int_{\xi_0}^{\xi_t}d\xi \cdot \bra{n,\xi} \nabla_\xi \ket{n,\xi} 
\, ,
\end{equation}
is the geometrical phase --- the latter form shows that it is time reparametrization invariant ($\nabla_\xi$ denotes the gradient in $\xi$-space). For a loop $C$ in $\xi$-space, Stokes' theorem may be used to cast~(\ref{gammandef}) in the form
\begin{equation}
\label{gamman2def}
\gamma_n(C)=\int_S d\xi^A d\xi^B \, K^{(n)}_{AB}
\, ,
\end{equation}
where 
\begin{equation}
\label{KABdef}
K^{(n)}_{AB}=i(\partial_A\bra{n,\xi})(\partial_B\ket{n,\xi})
\end{equation}
is the Berry curvature, and $S$ is any two-dimensional patch with $C$ as its boundary~\cite{Ber:84,Sim:83,Ber:88} (see also~\cite{Jac:89,Sha.Wil:89}).

When the initial state $\ket{n,t=0}$ is $d$-fold degenerate, the geometric phase factor $e^{i\gamma_n(t)}$ generalizes to a unitary matrix~\cite{Wil.Zee:84} (the Wilczek-Zee effect)
\begin{equation}
\label{Udef}
U_n(t)=P \exp \left(i\int_{\xi_0}^{\xi_t} d\xi \, A_n^\xi \right)
\, ,
\end{equation}
where $P \exp$ is a path-ordered exponential, and $A_n^\xi$ is a hermitean $d$ by $d$ matrix with entries 
\begin{equation}
\label{Amdef}
(A_n^\xi)_{ab}=i\bra{n,b;\xi} \nabla_\xi \ket{n,a;\xi}
\end{equation}
($a$, $b=1,\ldots ,d$, range over the degenerate subspace).
\section{Particles Constrained on Curves}
Consider a quantum particle constrained to move along a smooth curve in three dimensional euclidean space --- what confines the particle to the vicinity of the curve is a two-dimensional harmonic oscillator potential in the plane normal to the curve, of width $\eta$, with its minimum at the position of the curve. It is convenient to use an adapted coordinate frame, with coordinates $(s,\alpha,\beta)$, where $s$ is arclength along the curve, and $\eta \alpha$, $\eta \beta$ are distances along the normal $n$ and the binormal $b$ of the curve, respectively. Then the position vector $r(s,\alpha,\beta)$ of an arbitrary point in the vicinity of the curve is related to the position vector $R(s)$ of the curve itself by
\begin{equation}
\label{rsab}
r(s,\alpha,\beta)=R(s)+\eta \alpha n(s)+\eta \beta b(s)
\, .
\end{equation}
Taking $\eta \ll \kappa^{-1}$, guarantees that the frame is well defined in the region of physical interest --- we do assume that $\kappa(s) \neq 0$. The hamiltonian for the particle, expressed in terms of the adapted coordinates, is given by
\begin{equation}
\label{Hadapt}
H=-\frac{1}{2\sqrt{|G|}}\partial_A G^{AB} \sqrt{|G|}\partial_B+V(\alpha,\beta)
\, ,
\end{equation}
where $A$, $B$ range over the adapted coordinates, $G_{AB}$ is the metric induced from the ambient euclidean one, $G_{AB}=\partial_A r \cdot \partial_Br$, $G^{AB}$ its inverse, and $G=(1-\eta \alpha \kappa)^2$ its determinant~\footnote{We follow here the exposition in~\cite{Sch.Jaf:03} --- notice that the sign of $\eta$ in the expression for $G$ that follows from Eq.~(3) in that reference is opposite to the one given here --- we believe our expression is the correct one.}. For the confining potential we take $V(\alpha,\beta)=(\alpha^2+\beta^2)/2\eta^2$. Notice that $V$ only depends on the normal coordinates, so that, classically, the tangential motion of the particle is free. The normalization condition for the particle's wavefunction $\Phi$ is 
\begin{equation}
\label{Phinorm}
\int ds d\alpha d\beta \sqrt{|G|} \vert \Phi \vert^2=1
\, ,
\end{equation}
which motivates working with a rescaled wavefunction $\Psi=|G|^{1/4}\Phi$, obeying
\begin{equation}
\label{Psinorm}
\int ds d\alpha d\beta \vert \Psi \vert^2=1
\, ,
\end{equation}
so that $\int d\alpha d\beta \vert \Psi \vert^2$ can be interpreted as 
the probability density for finding the particle at the position $s$ along the curve. Accordingly, the hamiltonian undergoes a similarity transformation,
\begin{equation}
\label{Hsimtrans}
H \rightarrow \tilde{H}=\vert G \vert^{1/4} H \vert G \vert^{-1/4}
\, .
\end{equation}
Plugging in~(\ref{Hadapt}), and expanding in powers of $\eta$ results in
\begin{equation}
\label{Hres1}
\tilde{H}=\frac{1}{\eta^2} H_{-2}+H_0+\calO(\eta)
\, ,
\end{equation}
where
\begin{eqnarray}
\label{Hm2}
H_{-2}
\fe
-\frac{1}{2}
(\partial_\alpha^2+\partial_\beta^2)
+\frac{1}{2}(\alpha^2+\beta^2)
\ff
\label{H0}
H_0
\fe
-\frac{1}{2} (\partial_s-i \tau L)^2-\frac{\kappa^2}{8}
\, ,
\end{eqnarray}
and $L=i(\beta\partial_\alpha -\alpha \partial_\beta)$ is the generator of rotations in the normal plane. Looking for $\tilde{H}$ eigenstates, $\tilde{H}\Psi=\tilde{E}\Psi$, in the factorized form
\begin{equation}
\label{Psifact}
\Psi(s,\alpha,\beta)=\chi(\alpha,\beta) \psi(s)
\, ,
\end{equation}
one is led to consider simultaneous $H_{-2}$ and $L$ eigenkets $\chi^{(n)}_\sigma$, 
\begin{equation}
\label{chieigenHL}
H_{-2}\chi^{(n)}_\sigma=(n+1)\chi^{(n)}_\sigma
\, ,
\quad
L\chi^{(n)}_\sigma=\sigma \chi^{(n)}_\sigma
\, , 
\end{equation}
leading to
\begin{equation}
\label{Hpeigen}
-\frac{1}{2}\psi''_\sigma+i\sigma \tau \psi'_\sigma
+\frac{1}{2}
\left(
i\sigma \tau'+\sigma^2\tau^2-\frac{1}{4}\kappa^2
\right)\psi_\sigma=E_\sigma \psi_\sigma
\end{equation} 
for the tangential wavefunction
(primes denote derivatives with respect to $s$). Then $\Psi_\sigma=\chi^{(n)}_\sigma \psi_\sigma$ and $\tilde{E}=(n+1)/\eta^2+E_\sigma$.
\section{Wires With Quantum Memory}
Consider a particle constrained to move on a unit circle, then $\kappa(s)=1$ and $\tau(s)=0$. For the normal ket, take either of the degenerate doublet $\ket{\pm}=
(\ket{10}\pm i\ket{01})/\sqrt{2}$, with $H_{-2}$ eigenvalue 2, and  satisfying $L\ket{\pm}=\pm\ket{\pm}$ (we use the standard notation $\ket{10}\equiv a_\alpha^\dagger\ket{00}$, $\ket{01}\equiv a_\beta^\dagger\ket{00}$, where $a_\alpha^\dagger$, $a_\beta^\dagger$ are creation operators of excitations along the axes $\alpha$, $\beta$). Eq.~(\ref{Hpeigen}) then becomes
\begin{equation}
\label{Hpms}
-\frac{1}{2}\psi''_\sigma
-\frac{1}{8}\psi_\sigma=E_\sigma\psi_\sigma
\, ,
\end{equation}
with $\sigma=\pm 1$ and solutions
\begin{equation}
\label{sols}
\psi_\sigma^{(k)}
=
\frac{1}{\sqrt{2\pi}}e^{iks}
\, ,
\quad
E_\sigma^{(k)}=\frac{4k^2-1}{8}
\, ,
\quad 
k\in \mathbb{Z}
\, .
\end{equation}
We choose $k=0$ --- the corresponding 3D states $\Psi_\pm^{(0)}=\chi_\pm \psi_\pm^{(0)}$ form a degenerate doublet, with 
\begin{equation}
\label{chiE}
\chi_\pm=\rho e^{-\rho^2/2}e^{\pm i\phi}/\sqrt{\pi}
\, ,
\qquad 
\tilde{E}_\pm^{(0)}=2/\eta^2-1/8
\end{equation}
($\rho$, $\phi$ denote the standard polar coordinates in the normal plane).

Consider now a two-parameter deformation of the circle, given by
\begin{equation}
\label{cdeformCartesian}
\begin{split}
R(s;\xi,\zeta)
&=
(\cos s,\sin s,0)\\
& \quad {}+\xi(-\cos^3 s,\sin^3 s,0)+\zeta(0,0,\cos 2s)
\end{split}
\end{equation}
in cartesian coordinates $(x,y,z)$,
or
\begin{equation}
\label{cdeformAdapted}
\begin{split}
R(s;\xi,\zeta)
&=
(0,-1,0)\\
& \quad {}+\xi(\frac{1}{2}\sin 2s,\cos 2s,0)+\zeta(0,0,\cos 2s)
\end{split}
\end{equation}
in the adapted frame $(t,n,b)$,
with $\xi$, $\zeta \ll 1$ (in the notation of the previous section, $\xi^1 \equiv \xi$, $\xi^2 \equiv \zeta$). The corresponding velocity field $v=\partial R/\partial\xi|_{\xi=0}$ satisfies the {\em locally arclength preserving} condition $\partial_s v^t-\kappa v^n=0$ (similarly for $u=\partial R/\partial\zeta|_{\zeta=0}$). This enables the physical identification of points of the curve with the same $s$ coordinate, for different values of the deformation parameters, so that derivatives of the wavefunction with respect to the latter can be meaningfully taken. The above deformation brings along changes in $\kappa(s)$ and $\tau(s)$, which in turn result in the perturbation hamiltonian
\begin{equation}
\label{H1def}
H_1=\frac{3}{4} \xi \cos 2s +6i\zeta \sigma \left(
\sin 2s \,  \partial_s +\cos 2s \right) 
\, ,
\end{equation}
to be added to the zeroth-order hamiltonian in the left hand side of~(\ref{Hpms}). The resulting first order corrected ground state tangential wavefunctions are
\begin{equation}
\label{psi1def}
\psi_\pm=\frac{1}{\sqrt{2\pi}}\left(1-\big(\frac{3\xi}{8}\pm i 3\zeta\big)\cos 2s \right)
\, .
\end{equation}
We are still not in the position to use~(\ref{KABdef}) though. The reason is that, on the one hand, the true, physical wavefunction is the $\Phi$ that appears in~(\ref{Phinorm}), and not the rescaled $\Psi$ we have been working with, while, on the other hand, our expressions involve the adapted coordinates, which depend implicitly on the parameters $\xi$ and $\zeta$, since the curve being deformed in (\ref{cdeformCartesian}) drags with it the adapted frame. Our task then, in principle, would be to express the adapted coordinates in terms of the cartesian ones, differentiate them with respect to the parameters, treating the cartesian cordinates as constants, and reexpress the result in terms of the adapted coordinates. Then the derivatives with respect to the parameters in, \eg, (\ref{gammandef}), would contain contributions both from the explicit dependence of $\Phi(s,\alpha,\beta;\xi,\zeta)$ on the parameters, as well as the implicit one through the adapted coordinates. 
Taking the above into consideration, we find that
the matrix $A$ in~(\ref{Amdef}) is diagonal, so that the nondegenerate expression for $K$, Eq.~(\ref{KABdef}), may be used, giving
\begin{equation}
\label{Kpm}
K^{(0,\sigma)}_{\xi \zeta}=-\frac{9}{16}\sigma
\, .
\end{equation}
The fact that the two states in the doublet, corresponding to $\sigma=\pm 1$, pick up opposite geometrical phases, allows the detection of the latter by starting out the system in a suitable superposition, \eg, in the state $\ket{\Psi}_{t=0}=\ket{0,10}=(\ket{0,+}+\ket{0,-})/\sqrt{2}$, with wavefunction $\Psi\sim \alpha e^{-\rho^2/2}$ --- a plot of a constant probability density surface appears in Fig.~\ref{fig:tubeplot1}. Suppose now that the the systen is driven cyclically in the $\xi$-$\zeta$ plane, tracing the circle $\xi=\epsilon(1-\cos \lambda t)$, $\zeta=\epsilon \sin \lambda t$, with $\lambda \ll 1$, so that the adiabatic approximation is valid. For $t=0$ the wire described by~(\ref{cdeformCartesian}) is a unit circle in the $x$-$y$ plane, centered at the origin, with $n$ radially inwards, and $b$ along $z$. As $t$ increases, the wire sweeps out the self-intersecting surface shown in Fig.~\ref{fig:surface} (assuming $\epsilon=.2$). The geometrical phase accumulated after each revolution in parameter space by the state $\ket{0,\pm}$ is $\pm \Delta \phi$, where
\begin{equation}
\label{Dphirev}
\Delta \phi
= 
2\pi \epsilon^2 K^{(0,+)}_{\xi \zeta}
=
-\frac{9}{8} \pi \epsilon^2
\, ,
\end{equation}
so that after $m$ revolutions, the state vector is 
\begin{equation}
\label{ketmrev}
\ket{\Psi}_t
=
(e^{im\Delta \phi}\ket{0,+}+e^{-im\Delta \phi}\ket{0,-})/\sqrt{2}
\, .
\end{equation}
As a result, the corresponding probability density profile rotates in the plane normal to the curve by an angle $\Delta \phi$ --- Fig.~\ref{fig:tubeplot2} assumes an integer $m$ such that $\Delta \phi \approx \pi/2$.
\setlength{\figurewidth}{\columnwidth}
\begin{figure}
\begin{pspicture}(0\figurewidth,-.05\figurewidth)%
(\figurewidth,.58\figurewidth)
\setlength{\unitlength}{.25\figurewidth}
\psset{xunit=.25\figurewidth,yunit=.25\figurewidth,arrowsize=1.5pt
3}
\centerline{\raisebox{-.22\totalheight}{%
\includegraphics[angle=0,width=1.3\figurewidth]{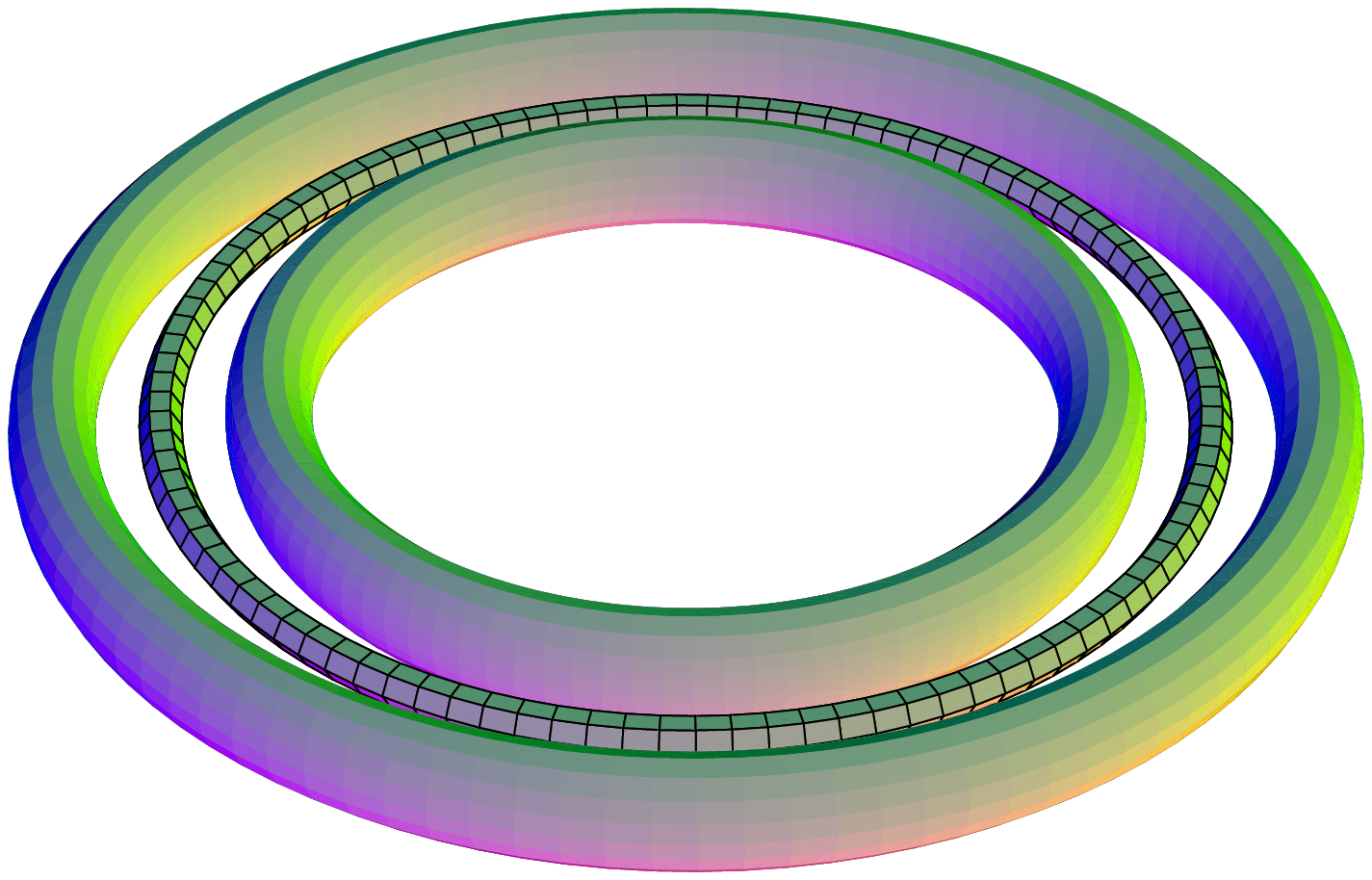}}}
\end{pspicture}
\capitem{\textsf{%
Plot of a constant probability density surface for the state $\ket{0,10}$, corresponding to excitation of the 2D harmonic oscillator along $n$. The darker colored ring represents the (undeformed) wire.%
}} 
\label{fig:tubeplot1}
\end{figure}
\setlength{\figurewidth}{\columnwidth}
\begin{figure}
\begin{pspicture}(0\figurewidth,-.05\figurewidth)%
(\figurewidth,.65\figurewidth)
\setlength{\unitlength}{.25\figurewidth}
\psset{xunit=.25\figurewidth,yunit=.25\figurewidth,arrowsize=1.5pt
3}
\centerline{\raisebox{-.22\totalheight}{%
\includegraphics[angle=0,width=1.3\figurewidth]{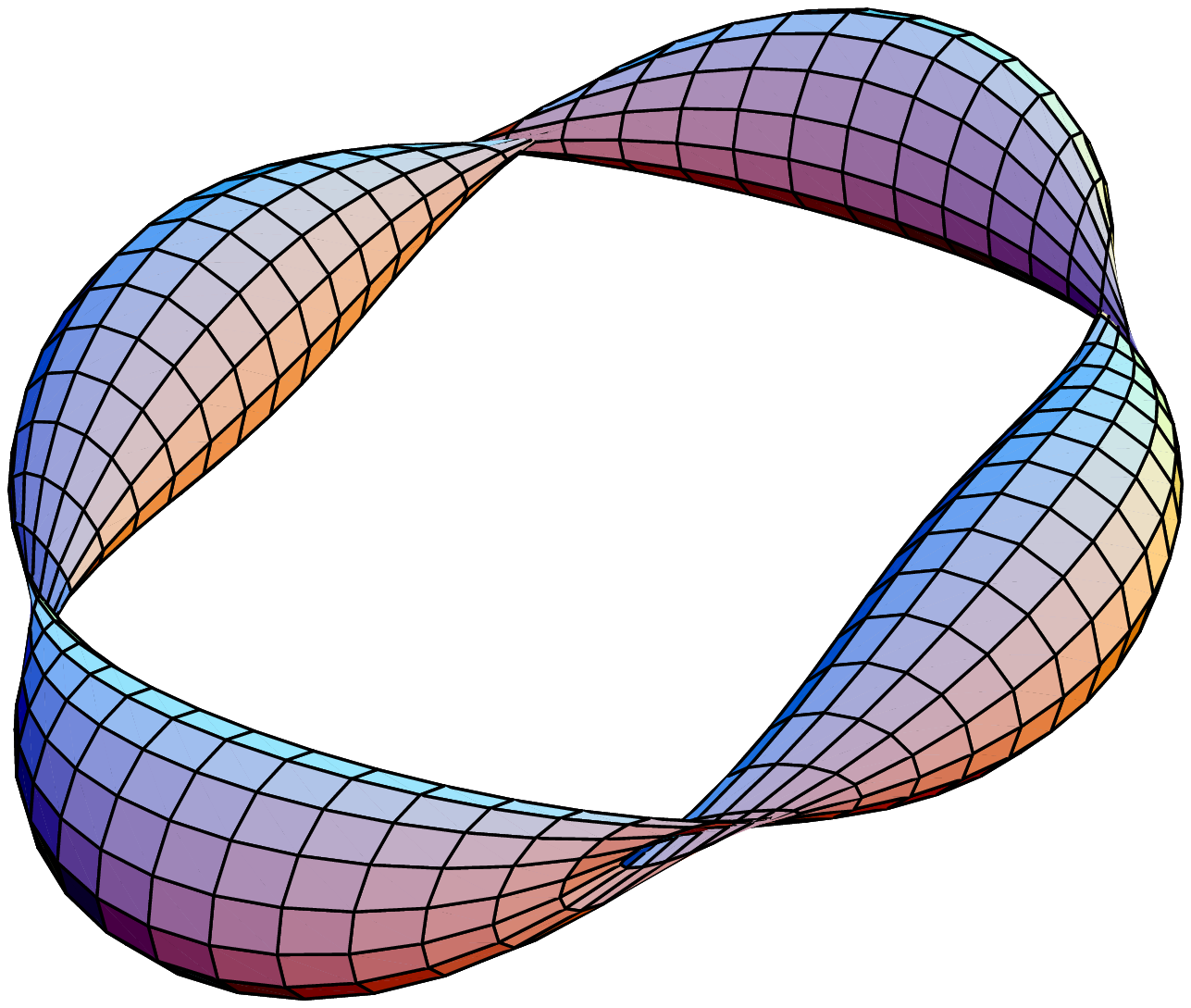}}}
\end{pspicture}
\capitem{\textsf{%
The surface swept out by the wire as it is being deformed, with $\xi$ and $\zeta$ tracing out a circle in parameter space.%
}} 
\label{fig:surface}
\end{figure}
\setlength{\figurewidth}{\columnwidth}
\begin{figure}
\begin{pspicture}(0\figurewidth,-.05\figurewidth)%
(\figurewidth,.58\figurewidth)
\setlength{\unitlength}{.25\figurewidth}
\psset{xunit=.25\figurewidth,yunit=.25\figurewidth,arrowsize=1.5pt
3}
\centerline{\raisebox{-.22\totalheight}{%
\includegraphics[angle=0,width=1.3\figurewidth]{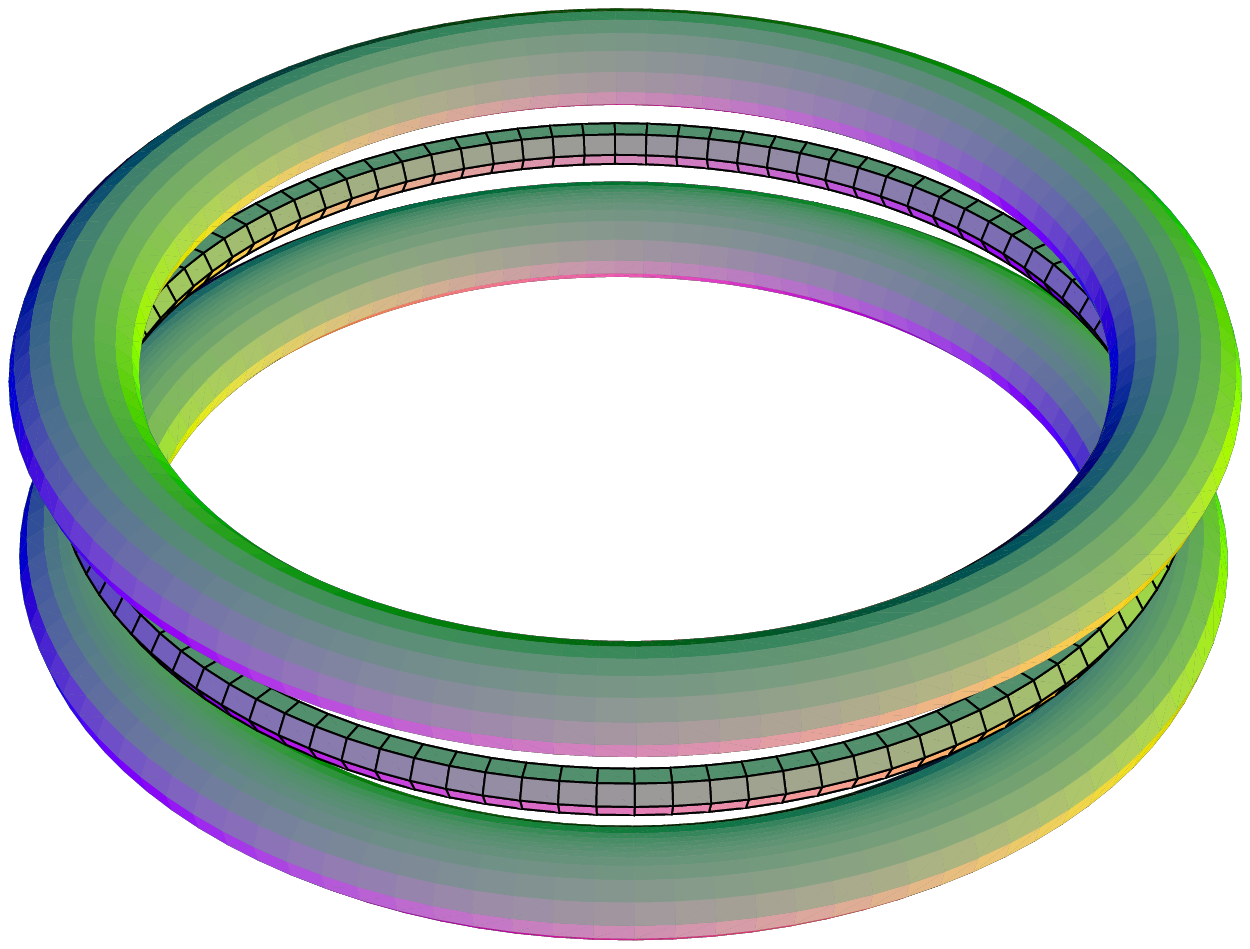}}}
\end{pspicture}
\capitem{\textsf{%
The accumulation of geometric phase is evidenced in the rotation of the probability density profile in the plane normal to the curve. Here it is assumed that $\Delta \phi \approx \pi/2$, so that the normal ket has changed from $\ket{10}$ (see  Fig.~\ref{fig:tubeplot1}) to $\ket{01}$.%
}} 
\label{fig:tubeplot2}
\end{figure}
\section{Conclusions}
We pointed out a wide class of quantum systems exhibiting geometrical phases: particles constrained to move along curves in three dimensional euclidean space, with the shape of the latter in the role of an infinite dimensional parameter space. We presented a particular example, involving deformations of a circle, and leading to a nonzero Berry curvature, Eq.~(\ref{Kpm}). The effect is observable, through changes in the probability density distribution, when superpositions of states are considered, each of which accumulates a different, in general, geometrical phase. 

There are obvious directions along which our result might be generalizable: higher dimension and/or codimension of the constraint manifold, \eg, the case of surfaces embedded in $\mathbb{R}^4$. Along with such endeavors, a detailed analytical study of the problem we considered here should provide general expressions for the curvature in terms of the initial shape of the wire and the velocity fields of the perturbations --- we defer such an analysis to a longer article, currently in progress. 

A question that we find particularly interesting is the extend to which the details of the deformation can be encoded in the probability density distribution. It is clear that, no matter what the initial quantum state of the particle is, there are deformations of the wire that pass unnoticed, \eg, those that evolve along an arbitrary curve in parameter space and then retrace their evolution along the same curve, back to the original shape. This means that one cannot, in general, reconstruct the deformation entirely from a comparison of the initial and final probability density distributions, but it is, nevertheless, plausible that, in general, more can be inferred than just the fact that there was {\em some} deformation.

Experimental uses of the above effect in quantum wires and nanotubes might be possible, as tiny periodic mechanical deformations are encoded {\em cumulatively} in a quantum system --- this could provide sensitive (periodic) motion detectors. In this context, it would be interesting to consider the potential amplification of the effect, either in the present, or its higher dimensional variants alluded to above, by replacing the single particle employed here by mesoscopic condensates. Another direction worth pursuing would be applications to holonomic quantum computing, 

C.{} C.{} would like to thank his colleagues at NTUA and, in particular, George Zoupanos and Konstantinos Anagnostopoulos, for hospitality, and multifaceted support. The authors acknowledge partial financial support from DGAPA-UNAM projects IN 121306-3 and IN 108103-3, as well as the EPEAEK programme ``Pythagoras II",
co-funded
by the European Union (75\%) and the Hellenic state (25\%).


\end{document}